\documentclass[twocolumn,aps,prl,superscriptaddress,amsmath,amssymb,superscriptaddress]{revtex4-2}

\usepackage{graphicx}
\usepackage{float}
\usepackage{natbib}
\usepackage{amsmath}
\graphicspath{ {./images/} }
\usepackage{xparse}
\usepackage{siunitx}
\usepackage{hyperref}
\usepackage{lipsum}
\usepackage{braket}

\begin{document}

\title{A quantum dot coupled to a suspended-beam mechanical resonator: from the unresolved- to the resolved-sideband regime}

\author{Clemens Spinnler}
\email{c.spinnler@unibas.ch}

\affiliation{Department of Physics, University of Basel, Klingelbergstrasse 82, CH-4056 Basel, Switzerland}

\author{Giang N. Nguyen}
\affiliation{Department of Physics, University of Basel, Klingelbergstrasse 82, CH-4056 Basel, Switzerland}

\author{Ying Wang}
\affiliation{Center for Hybrid Quantum Networks (Hy-Q), The Niels Bohr Institute, University of Copenhagen, DK-2100 Copenhagen Ø, Denmark}

\author{Marcel Erbe}
\affiliation{Department of Physics, University of Basel, Klingelbergstrasse 82, CH-4056 Basel, Switzerland}

\author{Alisa Javadi}
\altaffiliation[Current address: ]{School of Electrical and Computer Engineering, University of Oklahoma, Norman, OK 73019, USA}
\affiliation{Department of Physics, University of Basel, Klingelbergstrasse 82, CH-4056 Basel, Switzerland}

\author{Liang Zhai}
\altaffiliation[Current address: ]{Pritzker School of Molecular Engineering, University of Chicago, Chicago, IL 60637, USA}
\affiliation{Department of Physics, University of Basel, Klingelbergstrasse 82, CH-4056 Basel, Switzerland}

\author{Sven Scholz}
\affiliation{Lehrstuhl für Angewandte Festkörperphysik, Ruhr-Universität Bochum, DE-44780 Bochum, Germany}

\author{Andreas D. Wieck}
\affiliation{Lehrstuhl für Angewandte Festkörperphysik, Ruhr-Universität Bochum, DE-44780 Bochum, Germany}

\author{Arne Ludwig}
\affiliation{Lehrstuhl für Angewandte Festkörperphysik, Ruhr-Universität Bochum, DE-44780 Bochum, Germany}

\author{Peter Lodahl}
\affiliation{Center for Hybrid Quantum Networks (Hy-Q), The Niels Bohr Institute, University of Copenhagen, DK-2100 Copenhagen Ø, Denmark}

\author{Leonardo Midolo}
\affiliation{Center for Hybrid Quantum Networks (Hy-Q), The Niels Bohr Institute, University of Copenhagen, DK-2100 Copenhagen Ø, Denmark}

\author{Richard J. Warburton}
\affiliation{Department of Physics, University of Basel, Klingelbergstrasse 82, CH-4056 Basel, Switzerland}

\date{\today} 

\begin{abstract}
We present experiments in which self-assembled InAs quantum dots are coupled to a thin, suspended-beam GaAs resonator. The quantum dots are driven resonantly and the resonance fluorescence is detected. The narrow quantum-dot linewidths, just a factor of three larger than the transform limit, result in a high sensitivity to the mechanical motion. We show that one quantum dot couples to eight mechanical modes spanning a frequency range from 30 to \qty{600}{\MHz}: one quantum dot provides an extensive characterisation of the mechanical resonator. The coupling spans the unresolved-sideband to the resolved-sideband regimes. Finally, we present the first detection of thermally-driven phonon sidebands (at \qty{4.2}{\K}) in the resonance-fluoresence spectrum.
\end{abstract}

\maketitle
\begin{figure*}[t]
    \centering
    \includegraphics[width=0.95\textwidth]{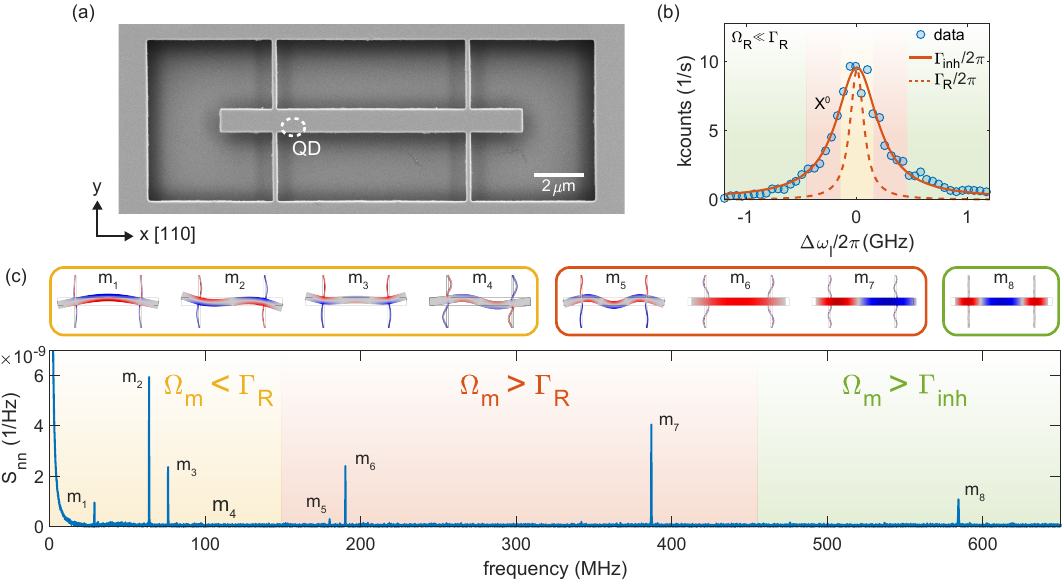}
    \caption{Mechanical resonator characterisation. (a) Scanning electron microscope image of the suspended-beam resonator. The designed resonator's size is $l = $ \qty{12.2}{\um} (length), $\mathrm{w} = $ \qty{960}{\nm} (width), $\mathrm{h} = $ \qty{180}{\nm} (height), $\mathrm{l_t} = $ \qty{180}{\nm} (tether length), and  $\mathrm{w_t} = $ \qty{2.8}{\um} (tether width). The quantum dot (QD) position is highlighted. (b) Resonant linewidth scan of the neutral exciton, $X^0$, of the QD highlighted in 
    (a). An inhomogeneously-broadened linewidth of $\Gamma_\mathrm{inh} = $ \qty{450}{\MHz} is extracted via a Lorentzian fit (orange line). The transform-limited linewidth is also plotted for which $\Gamma_\mathrm{R}/2\pi = $ \qty{150}{\MHz} (dashed orange line). (c) Noise-power spectrum of the thermal vibrations of the mechanical resonator obtained via a detuned autocorrelation measurement at \qty{4.2}{\K}. Displacement and exciton-phonon coupling profiles of the corresponding modes are shown on top, where the colour scheme represents the QD strain coupling. For (b) and (c): yellow, red, and green highlight the regions corresponding to the unresolved, the intermediate, and the resolved-sideband regime. }
    \label{fig:1}
\end{figure*}

Semiconductor quantum dots (QDs) are solid-state single-photon emitters. The transition energy of a QD reacts to strain such that the QD represents an ideal platform to investigate the interaction between a quantum emitter and mechanical motion \cite{Gell2008, Metcalfe2010, Nysten2017, Villa2017, Weiss2021, Wigger2021}. Specifically, strain leads to an energy shift of the QD's excited state (the exciton) via a deformation potential; the shift is mapped onto the emitted photons~\cite{Yeo2014}. The interaction between the QD and the mechanical system can be enhanced by phononic engineering, for example, via a mechanical resonator~\cite{Munsch2017,Montinaro2014}.\\
\indent To date, QDs have been coupled to various mechanical resonators with frequencies both below the QD's radiative decay rate (cantilevers, beams, trumpet resonators, and nanowires~\cite{Carter2018, Carter2019, Yuan2019, Vogele2020, Montinaro2014, Descamps2023, Yeo2014, Yoe2016, Munsch2017, Assis2017,Tumanov2018, Kettler2021, Finazzer2023}) and above the decay rate (surface-acoustic-wave cavities and phononic-crystal resonators~\cite{Imany2022,DeCrescent2022, spinnler_open_2023}).\\
\indent In this work, we couple self-assembled InAs quantum dots to a suspended-beam resonator. The QD's neutral exciton, $X^0$, is excited resonantly with a narrow-linewidth laser, and we detect the resonance fluorescence (RF). We present a complete characterisation of the QD properties (excited state lifetime, single-photon purity, and inhomogeneous linewidth broadening). We detect the thermal motion of the mechanical modes (at \qty{4.2}{\K}) and observe eight mechanical resonances from $29$ to \qty{584}{\MHz}. Further, we show that inhomogeneous broadening has to be included to make a distinction between unresolved- and resolved-sideband regimes. Finally, for the first time, we observe thermally-driven phonon sidebands in the RF spectrum, an important step towards heralded single-phonon creation/annihilation. \\
\indent The mechanical resonator consists of a \qty{180}{\nm}-thick semiconductor diode structure (see Ref.~\cite{spinnler_open_2023}). The diode consists of n- and p-doped layers for charge stabilisation and charge control of the QDs~\cite{Kuhlmann2013}. Figure~\ref{fig:1}(a) shows a scanning electron microscope image of the fabricated device. The tether length and attachment position are designed such that clamping losses should be minimal for the third in-plane breathing mode. The beam width is optimised such that the QD emission is maximised to the top (collection lens)~\cite{spinnler_open_2023}. The QDs are located in a plane at the centre of the membrane ($z=0$). The resonator contains about 50 QDs (see ~\hyperref[sec_Ap_QD]{Appendix C}) and we select a QD located close to the tether attachment point, displaced about halfway from the centre (in $y$-direction), Fig.~\ref{fig:1}(a).\\
\indent Figure~\ref{fig:1}(b) shows an optical linewidth scan with an excitation power much below saturation ($\Omega_\mathrm{R}\ll\Gamma_\mathrm{R}$, $\Omega_\mathrm{R}$ being the Rabi coupling, $\Gamma_\mathrm{R}$ the radiative decay rate). The transform-limited linewidth is $\Gamma_\mathrm{R}/2\pi = $ \qty{150}{\MHz} (see~\hyperref[sec_Ap_QD]{Appendix C}). The measured linewidth is larger, on account of noise in the semiconductor (mostly charge- but also spin-noise~\cite{Kuhlmann2013}), which results in an inhomogeneous broadening: $\Gamma_\mathrm{inh}/2\pi = $ \qty{450}{\MHz}. As the correlation times of the noise are much smaller than the measurement times in all the experiments presented here ~\cite{Kuhlmann2013, spinnler_open_2023}, inhomogeneous broadening needs to be considered in the interpretation of the measurements.\\ 
\indent The mechanical modes of the resonator are characterised via the Brownian motion (at \qty{4.2}{\K}), without any additional mechanical drive. We record an autocorrelation with a Hanbury Brown-Twiss setup~\cite{Munsch2017,spinnler_open_2023}. Since we wish to observe mechanical modes in the resolved-sideband regime, we need to work at elevated laser powers and increased laser detunings~\cite{spinnler_open_2023}, here, $\Omega_\mathrm{R} = 2\Gamma_\mathrm{R}$ and $\Delta\omega_\mathrm{l}/2\pi = $ \qty{300}{\MHz}. The noise-power spectrum, $S_\mathrm{nn}$, is obtained via a Fourier transform of the normalised autocorrelation, shown in Fig.~\ref{fig:1}(c). We observe eight peaks, each corresponding to a mechanical resonance. The modes are identified by comparing the frequencies and coupling strengths to finite-element simulations of the mechanics (see \hyperref[sec_Ap_comsol]{Appendix A}). Each mode, $\mathrm{m}_\mathrm{1}$ to $\mathrm{m}_\mathrm{8}$, corresponds to an in-plane mechanical resonance. There are five bending modes and three breathing modes. According to the simulations, the latter have a highly homogeneous coupling profile (in $y$-direction), facilitating the selection of a QD with large exciton-phonon coupling. The mechanical quality factor of the bending modes starts high at $Q_\mathrm{m_{1}} = 1.1\times10^4$, but reduces to $Q_\mathrm{m_{5}} = 8.2\times10^2$. We ascribe the increased damping to mode leakage to the surrounding substrate (clamping loss). Conversely, the breathing modes do not show a significant drop: $Q_\mathrm{m_{6}} = 3.8\times10^3$, $Q_\mathrm{m_{7}} = 3.9\times10^3$, and $Q_\mathrm{m_{8}} = 2.0\times10^3$. Although the design minimises clamping losses for $\mathrm{m}_\mathrm{8}$, $Q_\mathrm{m_{8}}$ is below $Q_\mathrm{m_{6}}$, and $Q_\mathrm{m_{7}}$. Thus, it is likely that the mechanical quality is limited by material-related losses, such as impurities at the surface or in the doping regions, possibly also Ohmic heating.\\
\indent In the absence of inhomogeneous broadening, the exciton-phonon interaction transitions from the unresolved- to the resolved-sideband regimes when the angular frequency of the mechanical resonator exceeds the QD's radiative decay rate. Here, this limit is not precise enough on account of significant inhomogeneous broadening (a major issue for samples without diode structure). We assign the mechanical modes to three different frequency regimes, see Fig.~\ref{fig:1}(c): (i) the unresolved-sideband regime with $\Gamma_\mathrm{R}>\Omega_\mathrm{m}$; (ii) the intermediate regime with $\Gamma_\mathrm{R}<\Omega_\mathrm{m}<\Gamma_\mathrm{inh}$; and (iii) the resolved-sideband regime with $\Omega_\mathrm{m}>\Gamma_\mathrm{inh}$. The three regimes are highlighted in yellow, red, and green in Fig.~\ref{fig:1}(b) and (d).\\
\begin{figure*}[ht]
    \centering
    \includegraphics[width=0.95\textwidth]{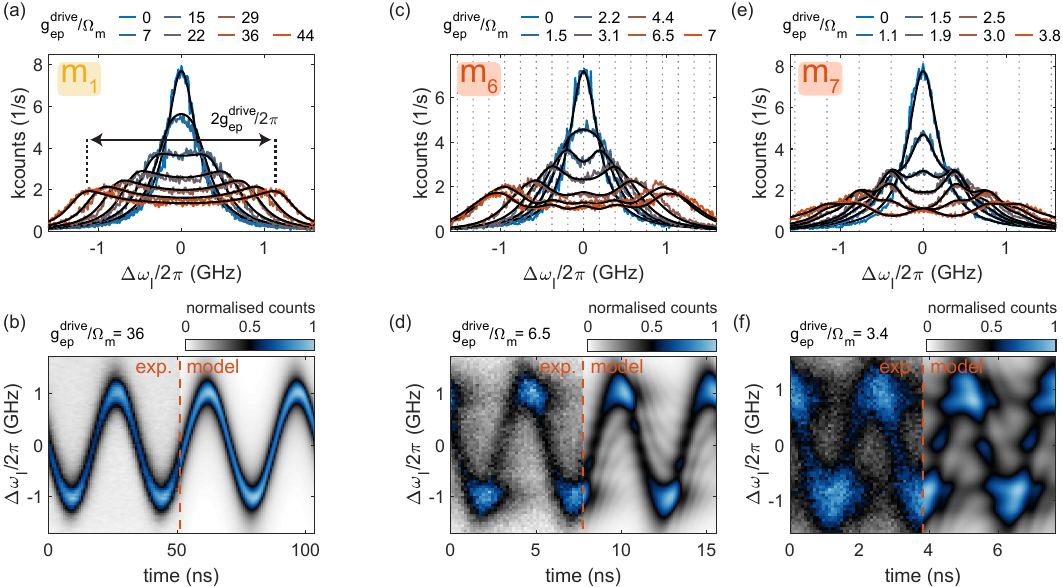}
    \caption{Mechanically modulated resonance fluorescence (RF) intensity. (a) Low-power, $\Omega_\mathrm{R}=0.1\Gamma_\mathrm{R}$, RF scans on driving of the first in-plane mechanical mode, $\mathrm{m}_1$. (b) Stroboscopic RF scan corresponding to (a). (c, e) Same as in (a) but on driving the first two in-plane breathing modes, $\mathrm{m}_6$ and $\mathrm{m}_7$, respectively. The vertical dashed lines highlight integer laser detunings of the mechanical frequency. (d, f) Stroboscopic RF scans corresponding to (c) and (e), respectively. The solid lines in (a), (c), (e) and the right panels in (b), (d), (f) show the results of the simulations. All parameters for the simulations are given in the text.}
    \label{fig:2}
\end{figure*}
\indent A straightforward way to compare the exciton-phonon interaction in the different regimes is to record a scan of RF intensity versus optical detuning. To enhance the interaction beyond Brownian motion, we drive the mechanical modes using an electric-field antenna, consisting of an open-ended coaxial cable mounted \qty{1}{\cm} above the sample surface~\cite{Huttel2010, spinnler_open_2023}. We find that the antenna couples strongly to the low-frequency modes, but not to $\mathrm{m_8}$, allowing us to perform mechanical driving in the unresolved and the intermediate regime. The optical excitation power is low such that power broadening is negligible (the coherent scattering regime, $\Omega_\mathrm{R}\ll\Gamma_\mathrm{R}$). We record two data sets. The first is the time-integrated RF scan, RF as a function of laser detuning. The second is a stroboscopic RF scan: the RF scans are recorded synchronised to the mechanical drive. The time tagger is synchronised with the mechanical drive and a time histogram is recorded for each laser detuning.\\
\indent Figure~\ref{fig:2}(a) and Fig.~\ref{fig:2}(b) show the time-integrated and stroboscopic RF scans in the unresolved-sideband regime, driving at $\Omega_\mathrm{m_1}/2\pi=~$ \qty{28.7}{\MHz}. The minimal frequency shift of the QD resonance is given by the frequency of a single phonon, $\Omega_\mathrm{m_1}/2\pi$, yet this is much smaller than the transform-limited linewidth. Thus, the QD resonance shifts linearly with the displacement of the driven mode, resulting in the sinusoidal modulation in the stroboscopic RF scan. In the time-integrated RF scan, the mechanical motion results in an increased linewidth, where the broadening directly corresponds to the QD frequency shift, $g_\mathrm{ep}^\mathrm{drive}/2\pi$. In comparison, Fig.~\ref{fig:2}(c) to (e) show the intermediate regime, driving at $\Omega_\mathrm{m_6}/2\pi=$~\qty{190}{\MHz} and $\Omega_\mathrm{m_7}/2\pi= $ \qty{387}{\MHz}. Here, the minimal shift of the QD resonance is larger than the transform-limited linewidth. Thus, side peaks are generated at multiple detunings of $\Omega_\mathrm{m}/2\pi$. However, the effect is washed out due to the inhomogeneous broadening. In the stroboscopic RF scans, Fig.~\ref{fig:2}(d) and (f), the side peaks can also be observed, with a highly reduced RF signal elsewhere. \\
\indent We model the interaction between the two systems, the driven two-level system (i.e., the QD) and the mechanical resonator, including the inhomogeneous broadening. We take a semi-classical master-equation approach~\cite{Weiss2021,Wigger2021}, where the mechanical motion couples dispersively to the excited state (the exciton) of the QD via $ \hat{H}_\mathrm{int} = \hbar g_\mathrm{ep}^\mathrm{drive}\sin{\left(\Omega_\mathrm{m}t\right)}\hat{\sigma}_\mathrm{+}\hat{\sigma}_\mathrm{-}$ (see~\hyperref[sec_Ap_model]{Appendix E}). Crucially, we include a \qty{300}{\MHz} frequency jitter of the QD resonance (Lorentzian distribution) to describe the inhomogeneous broadening. The model is plotted together with both time-integrated and stroboscopic RF scans. There is just one free parameter, the calibration factor which converts electrical power (applied to the antenna) to $g_\mathrm{ep}^\mathrm{drive}$. The model fits exceptionally well, Fig.~\ref{fig:2}(c) to (f). By linking the experimental results to the simulation, we find that the intensity of the side peaks depends on the ratio between the QD resonance shift and the mechanical frequency, $g_\mathrm{ep}^\mathrm{drive}/\Omega_\mathrm{m}$ \cite{Artioli2019,Weiss2021,DeCrescent2022}. Thus, the higher the mechanical frequency, the higher the mechanical driving needs to be in order to observe the same side-peak intensity.\\
\begin{figure*}[ht]
    \centering
    \includegraphics[width=0.95\textwidth]{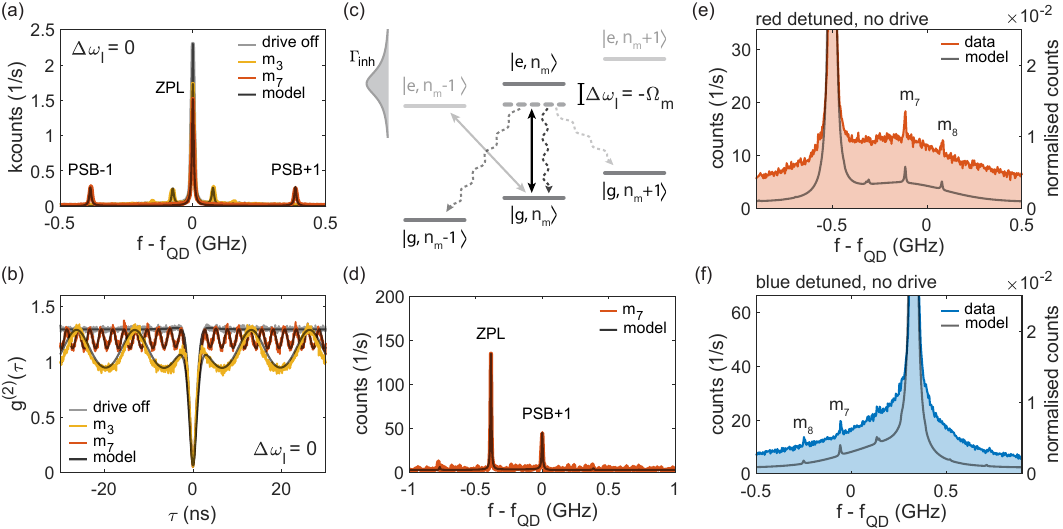}
    \caption{Mechanically modulated RF spectra in the coherent-scattering regime. (a) RF spectrum under driving of $\mathrm{m}_2$ and $\mathrm{m}_7$, with $\Omega_\mathrm{R}=0.2\Gamma_\mathrm{R}$. The zero-phonon line (ZPL) is accompanied by phonon sidebands (PSB), separated by $\pm \Omega_\mathrm{m}/2\pi$. The mechanical driving powers are $g_\mathrm{ep}^\mathrm{drive}/\Omega_\mathrm{m_3}= 1.3$ and $g_\mathrm{ep}^\mathrm{drive}/\Omega_\mathrm{m_7}= 0.65$, respectively. (b) Autocorrelation measurement of the RF using the same parameters as in (a). (c) Phonon-energy ladder diagram for the measurement shown in (d). The laser is detuned to the red sideband (grey double arrow) but still drives the elastic transition (black double arrow) due to inhomogeneous broadening ($\Gamma_\mathrm{inh}$). (d) Phonon sideband (PSB) of $\mathrm{m}_7$ under detuned optical driving at $\Delta\omega_\mathrm{l}=-\Omega_{\mathrm{m}_7}$ and $\Omega_\mathrm{R}=0.1\Gamma_\mathrm{R}$. The main contribution to the emission spectrum arises from elastic scattering (ZPL). The mechanical driving is $g_\mathrm{ep}^\mathrm{drive}/\Omega_\mathrm{m_7}= 0.43$. (e) and (f) RF spectrum at $\Omega_\mathrm{R}=0.1\Gamma_\mathrm{R}$ without an additional mechanical drive, i.e., measurement of the Brownian motion at \qty{4.2}{K}. To maximise the sideband intensity, the laser is red and blue detuned by \qty{-0.5}{\GHz} and \qty{+0.33}{\GHz}, respectively. The thermal phonon populations of $\mathrm{m_7}$ and $\mathrm{m_8}$ are 226 and 150, respectively }
    \label{fig:3}
\end{figure*}
\indent  Another way of investigating the interaction between the QD and the mechanical resonator is by recording the RF spectrum, i.e., RF intensity versus emission frequency for a constant laser frequency. 
To record the RF spectrum, we use a piezo-tunable Fabry-Perot etalon with a resolution of $\Delta\nu =  $ \qty{7.5}{\MHz} and a free spectral range of $\mathrm{FSR} = $ \qty{1.5}{\GHz}. We drive the QD weakly (i.e., in the coherent-scattering regime) and compare the RF spectra in the unresolved and in the intermediate regime. Figure~\ref{fig:3}(a) shows RF spectra upon driving $\mathrm{m}_3$ (yellow) and $\mathrm{m}_7$ (orange). The RF spectra contain a strong peak at the laser frequency (zero detuning in Figure~\ref{fig:3}(a)) and weak sidebands. The detunings of the side peaks correspond precisely to the mechanical frequencies. Hence, the peak at zero detuning corresponds to the zero-phonon line (elastic scattering); the side peaks to phonon sidebands (inelastic scattering). Both first-order and second-order sidebands are observed for $\mathrm{m}_3$.\\
\indent We stress that exploitation of the coherent scattering regime enables the phonon sidebands to be observed even though the system is not in the resolved-sideband regime. In the coherent scattering regime, a perfect two-level system produces single-photon replicas of the laser photons (laser linewidth is \qty{300}{\kHz}). Hence, the width of both the zero-phonon and phonon-sideband peaks in Fig.~\ref{fig:3}(a) are given by the linewidth of the Fabry-Perot etalon. Therefore, sidebands should be observed provided the mechanical frequency is larger than the filter resolution. A QD at low temperature shows weak spectrally broad RF on the red side of the main peak on account of inelastic scattering with longitudinal-acoustic (LA) phonons~\cite{Koong2019}. However, this background is at an insignificant level in Fig. \ref{fig:3}(a).\\
\indent We numerically model the RF spectra (see~\hyperref[sec_Ap_model]{Appendix E}), and the results are plotted on top of the data in Fig.~\ref{fig:3}(a). All the parameters are known at this point. Excellent agreement with the experimental data is achieved.\\
\indent To prove that the QD behaves as a single-photon emitter and that it interacts with single phonons, we perform autocorrelation measurements for the exact same measurement parameters as for the emission spectrum, shown in Fig.~\ref{fig:3}(b). The normalisation is performed via the detector count rates. The antibunching in the autocorrelation results in a single-photon purity of $95\%$. The antibunching in the emission's photon statistics has consequences for the interpretation of the sidebands in the RF spectra: the phonons must also be created and absorbed one-by-one~\cite{Weiss2021}. In essence, the quantum nature of the QD renders the classical input fields (laser, mechanical drive) into quantum outputs.\\
\indent The autocorrelation of the RF captures any time modulation created by the mechanical driving~\cite{Munsch2017}. In Fig.~\ref{fig:3}(b), we observe a strong time modulation with a frequency matching the mechanical driving frequencies, even for a laser detuning of zero. This is surprising at first sight since for zero laser detuning, the RF photon-flux from an ideal emitter is insensitive to first order to the mechanical motion (see~\hyperref[sec_Ap_model]{Appendix E}), and the second-order response would lead to a response at twice the mechanical frequency. The explanation lies in the inhomogeneous broadening. Due to the frequency jitter in the QD resonance, the laser spends a large amount of time off-resonance where the first-order sensitivity to mechanical motion is high, resulting in a modulation in the photon flux at the mechanical frequency. Furthermore, the inhomogeneous broadening reduces the ``width" of the anti-bunching dip by a factor of two, see~\hyperref[sec_Ap_model]{Appendix E}. All of this is well captured by our model, plotted on top of the data in Fig.~\ref{fig:3}(b).\\
\indent Figure~\ref{fig:3}(d) shows the RF spectrum with the pump laser tuned to the red sideband of $\mathrm{m}_7$. This enhances the red sideband at the expense of the blue sideband. For an ideal two-level system, a signature of the resolved-sideband regime is that the phonon sideband dominates over elastic scattering~\cite{Wigger2021} (i.e. the right-hand peak would dominate in Fig.~\ref{fig:3}(d), see~\hyperref[sec_Ap_model]{Appendix E}). This is not achieved for $\mathrm{m}_7$ even though $\Omega_\mathrm{m}>\Gamma_\mathrm{R}$ (see~\hyperref[sec_Ap_model]{Appendix E}). The explanation lies in the frequency jitter of the QD exciton. Fig.~\ref{fig:3}(c) shows the phonon ladder with the laser tuned to the red sideband, $\ket{g,n_\mathrm{m}} \leftrightarrow\ket{e,n_\mathrm{m}-1}$. The laser still drives the direct transition $\ket{g,n_\mathrm{m}} \leftrightarrow\ket{e,n_\mathrm{m}}$ but in a detuned manner. Frequency jitter of the QD resonance tends to increase coherent scattering from the direct transition and reduce coherent scattering from the phonon sideband. Simulations including the frequency jitter reproduce the experimental results convincingly, Fig.~\ref{fig:3}(d).\\
\indent The ultimate goal is to operate QD-mechanical-resonator devices in the single-phonon regime~\cite{DeCrescent2022}. As a first step, we present measurements of the phonon sidebands relying only on thermal phonons at \qty{4.2}{\K}. This constitutes a measurement of the Brownian motion. We work in the coherent scattering regime and to maximise the sideband intensity, the laser is detuned from the QD exciton. Figure~\ref{fig:3}(e) and (f) show the RF spectra for red- and blue-detuned pumping, respectively. We observe two sidebands corresponding to the highest two breathing modes, $\mathrm{m}_7$ and $\mathrm{m}_8$, with a thermal phonon occupation of 226 and 150, respectively. The intensity of the sidebands, with respect to the peak at the laser frequency, is around $10^{-3}$. This matches well with the expectations from our model, for which we take the thermal coupling rate from finite-element simulations (see~\hyperref[sec_Ap_comsol]{Appendix A}). Also here, the sideband intensity depends on $g_\mathrm{ep}^\mathrm{th}/\Omega_\mathrm{m}$, meaning that both need to be increased equally, to observe sidebands at higher mechanical frequencies. In addition to the narrow peaks, there is a broad background. This arises from scattering between LA phonons and the exciton~\cite{Koong2019}: the phonon resonances $\mathrm{m}_\mathrm{1}$ to $\mathrm{m}_\mathrm{8}$ sit on a broad background in the phonon density-of-states function. (We note that these broadband phonons, a few hundred GHz, are not included in the model; also, that scattering with these phonons is imperfectly filtered in the experiment due to the small $\mathrm{FSR}$, such that the background in Fig.~\ref{fig:3}(e) and (f) is about an order-of-magnitude higher than the real one.)\\
\indent In summary, we present a GaAs mechanical resonator, a suspended beam, and show that a single QD couples to eight mechanical resonances in the frequency range 30 to \qty{600}{\MHz}. This allows us to probe the coupling spanning the unresolved- to the resolved-sideband regimes with just one QD and one mechanical device. We find that:

(i) The breathing modes result in high mechanical resonance frequencies even though the resonator itself has microscopic (and not nanoscopic) dimensions.

(ii) Charge control of the QD is preserved in the mechanical resonator and the optical linewidth is close to the transform limit.

(iii) Finite-element simulations reproduce the measured mechanical frequencies and coupling strengths.

(iv) The coupling rates are high such that thermally-driven phonon sidebands in the QD resonance fluorescence are observed.

(v) The coupling to the 8th mode is in the resolved sideband regime: the mechanical frequency is larger than the total frequency-width of the QD exciton.


\section*{Acknowledgements}
We thank Lukas Sponfeldner for the SEM images. We also thank Martino Poggio, Tomasz A.\ Jakubczyk, Yannik Laurent Fontana, Hinrich Mattiat, and Thibaud Ruelle for stimulating discussions. \\
\indent The work was supported by SNF Project 200020\_204069. GNN, ME, LZ, and AJ received funding from the European Union’s Horizon 2020 Research and Innovation Programme under the Marie Sk\l{}odowska-Curie grant agreement No.\ 861097 (QUDOT-TECH), No.\ 721394 (4PHOTON),  and No.\ 840453 (HiFig), respectively. YW, PL and LM acknowledge financial support from Danmarks Grundforskningsfond (DNRF 139, Hy-Q Center for Hybrid Quantum Networks). LM acknowledges the European Research Council (ERC) under the European Union’s Horizon 2020 research and innovation programme (Grant Agreement No. 949043, NANOMEQ). SS, ADW and AL acknowledge financial support from the grants DFH/UFA CDFA05-06, DFG TRR160, DFG project 383065199, and BMBF Q.Link.X 16KIS0867.

\begin{table*}[htp]
\caption{Measurement and simulation parameters of the mechanical modes. The first two rows show the measured mechanical resonance frequencies, $\Omega_\mathrm{m}/2\pi$, and mechanical quality factors, $Q_\mathrm{m}$. The remaining rows show simulated parameters (taking the QD used experimentally): mechanical eigenfrequency, $\Omega^{\mathrm{sim}}_\mathrm{m}/2\pi$, effective mass, $m_{\mathrm{eff}}$, spring constant, $k$, vacuum fluctuation, $x_{\mathrm{zpf}}$, thermal displacement at \qty{4.2}{\K}, $x_{\mathrm{th}}$, thermal phonon occupation, $n_\mathrm{m}$, exciton-phonon coupling rate, $g_\mathrm{ep}$, and thermally enhanced coupling rate, $g^\mathrm{th}_\mathrm{ep}$, obtained from finite-element simulations.}
\vspace{0.3 cm}
\centering
\begin{tabular}{l|cccccccc}\hline\hline
mode number & 1 & 2 & 3 & 4 & 5 & 6 & 7 & 8 \\
\hline
$\Omega_\mathrm{m}/2\pi$ (MHz) & 28.7 & 63.9 & 76.1 & 112 & 180 & 190 & 387 & 584 \\
$Q_\mathrm{m}$ & $1.05\times10^4$ & $1.11\times10^4$ & $1.60\times10^3$ & $2.70\times10^3$ & $8.15\times10^2$ & $3.81\times10^3$ & $3.94\times10^3$ & $2.01\times10^3$ \\
\hline
$\Omega^{\mathrm{sim}}_\mathrm{m}/2\pi$ (MHz) & 29 & 68 & 80 & 114 & 185 & 194 & 396 & 596 \\
$m_{\mathrm{eff}}$ ($10^{-15}$ kg) & 4.8 & 1.6 & 2.0 & 5.0 & 5.2 & 6.2 & 5.7 & 5.6 \\
$k$ ($10^{3}$ N/m) & 0.16 & 0.30 & 0.52 & 2.5 & 6.9 & 9.2 & 35 & 78 \\
$x_{\mathrm{zpf}}$ ($10^{-15}$ m) & 7.7 & 8.7 & 7.0 & 3.8 & 2.9 & 2.6 & 1.9 & 1.6 \\
$x_{\mathrm{th}}$ ($10^{-13}$ m) & 5.9 & 4.3 & 3.2 & 1.4 & 0.9 & 0.8 & 0.4 & 0.3\\
$\langle n_\mathrm{m}\rangle$ & 3050 & 1370 & 1150 & 781 & 486 & 460 & 226 & 150\\
$g_\mathrm{ep}/2\pi$ (kHz) & 50 &   216 &   208  &  7  &  223  &  318  &  693 &   737\\
$g^\mathrm{th}_\mathrm{ep}/2\pi$ (MHz) & 3.8 &  10.9  &  9.5  &  0.3   & 6.7  &  9.5 &  14.6 &  12.6\\ \hline\hline
\end{tabular}
\label{table:FB_comsol}
\end{table*}

\section*{Appendix A: Finite-element simulations}
\label{sec_Ap_comsol}
Since the displacement profile of the mechanical modes is non-trivial, we simulate the mechanical resonator using Comsol Multiphysics. We perform an eigenmode study where we apply fixed boundary conditions to the ends of the four tethers. We also apply a symmetry condition with respect to $z=0$ to obtain only in-plane mechanical modes. The mechanical resonator is described by a harmonic oscillator via:  
\begin{equation}
u(r,t) =x(t)\lvert u(r)\rvert,
\end{equation}
where $\lvert u(r)\rvert = \frac{u(r)}{\mathrm{max}(\lvert u(r) \rvert)}$ is the normalised displacement~\cite{Hauer2013} obtained from COMSOL. $x(t)$ is the time-dependent displacement defined by the equation of motion:  
\begin{equation}
m_\mathrm{eff}\frac{dx^2(t)}{dt^2}+m_\mathrm{eff}\Gamma_\mathrm{m}\frac{dx(t)}{dt}+m_\mathrm{eff}\Omega_\mathrm{m}^2
x(t) = F(t),
\end{equation}
where $\Omega_\mathrm{m}/2\pi$ is the mechanical frequency, $k = m_\mathrm{eff}\Omega_\mathrm{m}^2$ the spring constant, and $\Gamma_\mathrm{m}$ the energy dissipation rate, related to the mechanical quality by $Q_\mathrm{m} = \Omega_\mathrm{m}/\Gamma_\mathrm{m}$. \\
\indent We estimate the exciton-phonon coupling rate via the effective mass and zero-point motion, obtained via thermomechanical calibration~\cite{Hauer2013} by performing a volume integration of the normalised displacement:
\begin{equation}
    m_{\mathrm{eff}} = \int \rho  \left( \frac{u(r)^2}{\mathrm{max}(\lvert u(r) \rvert)^2} \right) \,dV,
\end{equation}
\begin{equation}
    x_{\mathrm{zpf}} = \sqrt{\frac{\hbar}{2m_{\mathrm{eff}}\Omega_\mathrm{m}} },
\end{equation}
where $\rho$ is the material density of GaAs, and $\hbar$ is the reduced Planck constant. The exciton-phonon coupling rate $g_\mathrm{ep}$ is: 
\begin{equation}
    g_\mathrm{ep} = \frac{\partial \omega_\mathrm{QD}}{\partial x}x_{\mathrm{zpf}}=\frac{\Delta E}{\hbar}, 
    \label{eq_8}
\end{equation}
where $\Delta E$ describes the deformation potential coupling. $\Delta E$ is obtained from the strain profile after normalising the displacement to the zero-point motion and reads~\cite{Munsch2017,Yeo2014}:   
\begin{equation}
    \Delta E = a(\epsilon_{xx}+\epsilon_{yy}+\epsilon_{zz})-\frac{b}{2}(\epsilon_{xx}+\epsilon_{yy}-2\epsilon_{zz}),
\end{equation}
where $\epsilon_\mathrm{ii}$ are the strain componends, and $a=-8.33~e\mathrm{V}$ and $b=-1.7~e\mathrm{V}$ are the deformation potentials for the hydrostatic and shear strain of GaAs, respectively~\cite{Vurgaftman2001,Walle1989}.\\
\indent The thermally enhanced coupling rate, $g_\mathrm{ep}^\mathrm{th}$, is obtained via the thermal displacement~\cite{Leijssen2017}:
\begin{equation}
    g_\mathrm{ep}^\mathrm{th} = \frac{g_\mathrm{ep}}{x_\mathrm{zpf}}x_\mathrm{th}.
\end{equation}
The thermal displacement, $x_{\mathrm{th}}$, is given by the equipartition theorem~\cite{Yeo2014}:
\begin{equation}
    x_{\mathrm{th}} = x_{\mathrm{zpf}}\sqrt{\frac{2k_\mathrm{B}T}{\hbar \Omega_\mathrm{m}}} =  x_\mathrm{zpf}\sqrt{2\langle n_\mathrm{m}\rangle},
\end{equation}
where $T$ is the phonon-bath temperature, $\langle n_\mathrm{m}\rangle$ the phonon occupation of the corresponding mechanical mode, and $k_\mathrm{B}$ the Boltzmann constant. \\
\indent Table~\ref{table:FB_comsol} shows the relevant parameters of all mechanical modes, obtained via the finite-element simulation, along with the measured mechanical frequencies and quality factors, obtained from the noise-power measurements, see~\hyperref[sec_Ap_mech_Q]{Appendix D}.

\section*{Appendix B: Measurement setup and methods}
\label{sec_Ap_setup}
Preliminary characterisation is carried out by exciting the QDs with an above-bandgap laser, collecting the photoluminescence (PL). All measurements of the exciton-photon coupling rely on resonant excitation, exciting with a laser in resonance with the QD, collecting the resonance fluorescence (RF). The power of the resonant laser is stabilised with a double-pass acoustic-optic modulator setup and the frequency is stabilised using a wavemeter. The lasers are fibre-coupled and sent to the He-bath cryostat. The sample is mounted on an x/y/z-piezo stack in a home-built measurement tube, and we excite and collect from the top via a lens with NA = 0.65. The focal spots are diffraction limited. We use a darkfield microscope~\cite{Kuhlmann2013a} that suppresses the reflected laser light using two polarising beam-splitters in combination with a linear polariser and a quarter-wave plate. The microscope head also includes a camera for imaging of the sample surface. The collected QD photons are fibre-coupled and sent to the detection setup. We use either a spectrometer with CCD detector-array or superconducting nanowire single-photon detectors (SNSPDs) in combination with a Hanbury Brown-Twiss setup and a Fabry-Perot etalon. \\
\indent In all our measurements, we take special care to suppress the reflected laser light and to correct for drifts in the QD resonance frequency. Every minute, the laser suppression is automatically adjusted and the QD is tuned to the laser frequency by performing a fast linewidth scan, tuning the QD frequency via the gate voltage. When using the Fabry-Perot etalon as filter, a 90:10 fibre beam-splitter is installed in front of the etalon such that $10\%$ of the unfiltered QD photons can be used to perform the locking and suppression task. The etalon itself is also stabilised actively by locking it to the laser frequency. During the stroboscopic RF scans, mechanical phase drifts are observed due to heating by the electric-field antenna. The full 2D scan (time and frequency axes) is performed within a minute and then repeated many times. The phase is corrected in a post-processing step~\cite{spinnler_open_2023}. For details on the fabrication of the mechanical resonator see Ref.~\cite{Uppu2020}.
\vspace{0.7cm}

\section*{Appendix C: QD characterisation}
\label{sec_Ap_QD}
\begin{figure}[t]
\includegraphics{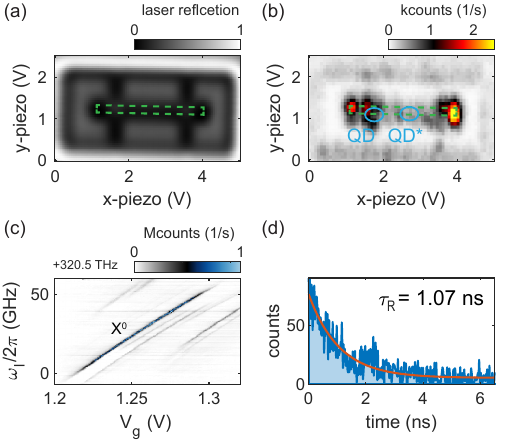}
\caption{\label{fig:1s} QD characterisation. (a) Laser reflection map of the mechanical device, recorded using an x/y-piezo scanner. The beam position is highlighted in green. (b) Photoluminescence map, recorded for the same area as in (a). The positions of the beam and of the QDs are highlighted. (c) Resonance fluorescence charge plateau map of the neutral exciton of the QD from the main text. The QD shows a fine structure splitting of almost \qty{10}{\GHz}. (d) Radiative lifetime measurement with $\tau_\mathrm{R} = $ \qty{1.07}{\ns}, resulting in $\Gamma_\mathrm{R}=2\pi\times $ \qty{150}{\MHz}.}
\end{figure}
To select a QD we perform a laser-reflection map and a PL map, shown in Fig.~\ref{fig:1s}(a) and (b), respectively. For the laser-reflection map, the microscope head is in brightfield mode and the reflected laser is measured with a standard photodiode. The PL map is carried out by recording a full PL-spectrum for every position and post-processing the data in a given wavelength range, here $930-960~\mathrm{nm}$. To detect all of the mechanical modes, we perform measurements on two quantum dots: the QD from the main text located close to the tethers, and QD$^*$ located in the centre of the beam (data in~\hyperref[sec_Ap_mech_Q]{Appendix D}). The beam and the QD positions are highlighted in Fig.~\ref{fig:1s}. Note that the above-band laser can introduce additional mechanical losses~\cite{spinnler_open_2023}. \\
\indent We perform our measurements on the neutral exciton, $X^0$. For all measurements, the Rabi coupling is calibrated with a power-series measurement. Figure~\ref{fig:1s}(c) shows a low-power RF-versus-voltage scan of the QD from the main text. The measurement is performed by scanning the gate voltage, $V_g$, for every laser frequency and counting the emitted photons with an SNSPD. Figure~\ref{fig:1}(b) shows an RF scan in the centre of the charge plateau. The scan is well described by a Lorentzian function with a linewidth of $\Gamma_\mathrm{inh}/2\pi= $ \qty{450}{\MHz}. To estimate the transform-limited linewidth, we perform a lifetime measurement. Figure~\ref{fig:1s}(d) shows a time histogram of a rapid-adiabatic-passage measurement: $\mathrm{m}_1$ is driven strongly ($g_\mathrm{ep}^\mathrm{th}/2\pi = $ \qty{60}{\GHz}) such that the QD passes the laser within \qty{200}{\ps}~\cite{spinnler_open_2023}. An excited state lifetime of $\tau_\mathrm{R}= $ \qty{1.07}{\ns} is extracted via an exponential fit. The corresponding transform-limited linewidth is $\Gamma_\mathrm{R}/2\pi = $ \qty{150}{\MHz}. Therefore, the inhomogeneous contribution to the linewidth is \qty{300}{\MHz}, just twice the homogeneous contribution. Equivalently, the measured linewidth is only a factor of three above the transform limit.    
\begin{figure}[t]
\includegraphics{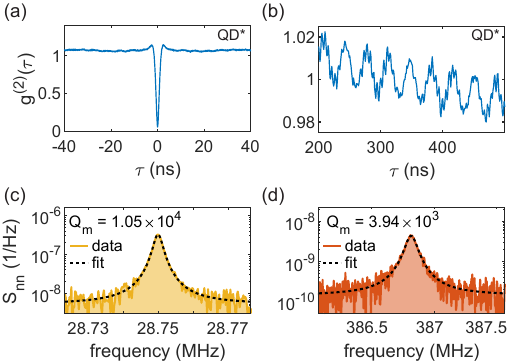}
\caption{\label{fig:2s} Mechanical characterisation. (a) Autocorrelation of the resonance fluorescence from QD$^*$ showing an antibunching dip. (b) Zoom-in at long-time delays of the autocorrelation shown in (a). Several mechanical oscillations are visible. (c) and (d) Noise-power spectra of $\mathrm{m}_1$ and $\mathrm{m}_7$ obtained via a Fourier transform of high-resolution and long-time-delay autocorrelation measurements. To extract the mechanical quality factors, the thermal noise peaks are fitted with Lorentzian functions including a y-offset.}
\end{figure}

\section*{Appendix D: Mechanical characterisation} 
\label{sec_Ap_mech_Q}
The mechanical resonator is characterised by measuring the autocorrelation function, $g^{(2)}(\tau)$, of the QD's photons. A detailed description of which laser power and laser detuning should be chosen can be found in Refs.~\cite{Munsch2017, spinnler_open_2023}. Figure~\ref{fig:2s}(a) shows an autocorrelation measurement of QD$^*$. The low $g^{(2)}(0) = 0.05$ is proof of single-photon emission. The residual $g^{(2)}(0)$ is a consequence of a small amount of laser light leaking into the collection channel. Clear mechanical oscillations are visible in $g^{(2)}(\tau)$, see Fig.~\ref{fig:2s}(b). The strong oscillation arises from the fundamental bending mode, while the weaker oscillations arise from the breathing modes. The noise-power spectrum is obtained from the autocorrelation data via a Fourier transform~\cite{Munsch2017}, exploiting the Wiener–Khinchin theorem: 
\begin{equation}
    S_{\mathrm{nn}}(f) = 2\mathrm{FFT}\left[g^{(2)}(\tau)\right]\tau_{\mathrm{bin}},
    \label{eq:FFT}
\end{equation}
where $g^{(2)}(\tau)$ is the normalised autocorrelation, and $\tau_\mathrm{bin}$ is the binning time. No background correction is performed for the noise-power spectrum and the noise floor is given by~\cite{spinnler_open_2023}: 
\begin{equation}
    S_\mathrm{nn}^\mathrm{noise-floor} = \frac{2}{\langle n \rangle\sqrt{ t_\mathrm{int}/t_\mathrm{del}}},
\end{equation}
where $\langle n \rangle$ is the average QD count rate, $t_\mathrm{int}$ is the measurement time, and $t_\mathrm{del}$ is the maximum delay in the autocorrelation measurement.\\
\indent We extract the resonance frequencies and quality factors of the mechanical modes directly from the noise-power spectrum. Figure~\ref{fig:2s}(c) and (d) show high-resolution spectra of $\mathrm{m}_1$ and $\mathrm{m}_7$ fitted to Lorentzian functions. A mechanical quality of $Q_\mathrm{m_1} = 1.05\times10^4$ and $Q_\mathrm{m_7} = 3.81\times10^3$ is extracted, respectively. Details on the remaining mechanical-noise peaks can be found in Tab.~\ref{table:FB_comsol}. 

\section*{Appendix E: Master-equation model}
\label{sec_Ap_model}
\begin{figure}[t]
\includegraphics{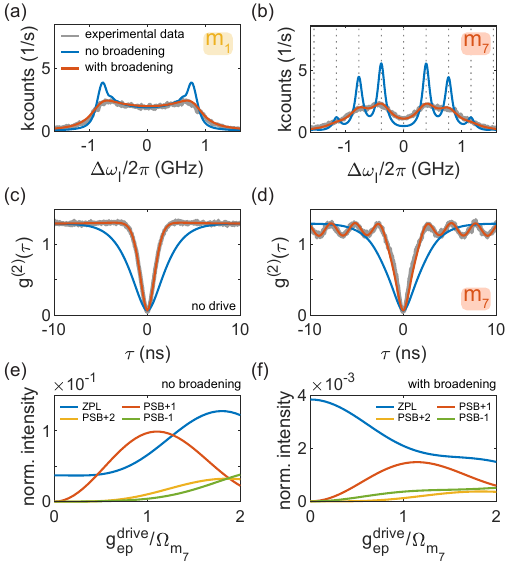}
\caption{\label{fig:3s} Master-equation simulations. (a) and (b) Resonance fluorescence intensity scans under driving of $\mathrm{m}_1$ and $\mathrm{m}_7$, respectively. (c) Autocorrelation measurements of the resonance fluorescence without mechanical driving, and (d) with driving of $\mathrm{m}_7$. In (a) to (d) the simulations are plotted with (red) and without (blue) inhomogeneous broadening. (e) and (f) Simulations of the intensity of zero-phonon line (ZPL) and phonon sidebands (PSB) under driving of $\mathrm{m}_7$, without and with inhomogeneous broadening, respectively. The laser is detuned to $\Delta\omega_\mathrm{l} = -\Omega_\mathrm{m_7}$. The intensities are normalised to the ZPL intensity for zero detuning, zero drive, and no broadening. }
\end{figure}

We model the interaction between the QD and the mechanical resonator in a semi-classical way. The bare Hamiltonian reads: 
\begin{equation}
\hat{H}= \hbar\omega_\mathrm{QD}\hat{\sigma}_\mathrm{+}\hat{\sigma}_\mathrm{-} + \hbar\Omega_\mathrm{m}\left( \hat{b}^\dagger  \hat{b} +1/2\right),
\end{equation}
where the first term is the energy of the QD and the second term is the energy of the mechanical resonator, described by a quantum harmonic oscillator. $\langle n_\mathrm{m}\rangle = \langle\hat{b}^\dag\hat{b}\rangle$ is the phonon population of the mechanical resonator, and $\hat{\sigma}_\mathrm{+} = \ket{e}\bra{g}$ and $\hat{\sigma}_\mathrm{-} = \ket{g}\bra{e}$ are the transition operators. The interaction of the laser with the QD in the dipole approximation reads: 
\begin{equation}
\hat{H}_\mathrm{drive} = -\hbar\frac{\Omega_\mathrm{R}}{2}\left( \hat{\sigma}_\mathrm{+}+\hat{\sigma}_\mathrm{-}\right) \left( e^{i\omega_\mathrm{l}t}+e^{-i\omega_\mathrm{l}t}\right),
\end{equation}
where $\Omega_\mathrm{R}$ is the optical Rabi coupling, and $\omega_\mathrm{l}/2\pi$ is the frequency of the driving laser. The mechanical motion couples to the excited state of the QD, the exciton, and is described in the dispersive regime. We assume that the mechanical phase is static on the time dynamics of the QD~\cite{Weiss2021}: 
\begin{equation}
    \hat{H}_\mathrm{int} = \hbar g_\mathrm{ep}^\mathrm{th}\sin{\left(\Omega_\mathrm{m}t\right)}\hat{\sigma}_\mathrm{+}\hat{\sigma}_\mathrm{-} ,
\end{equation}
where $g_\mathrm{ep}^\mathrm{th}=g_\mathrm{ep}\sqrt{2\langle n_\mathrm{m}\rangle}$ is the thermally-enhanced exciton-phonon coupling rate~\cite{Leijssen2017} and $ \hat{\sigma}_\mathrm{+}\hat{\sigma}_\mathrm{-} = \ket{e}\bra{e}$. The radiative decay of the QD exciton is added via a collapse operator $\hat{L}= \sqrt{\Gamma_\mathrm{R}}\hat{\sigma}_\mathrm{-}$. Finally, the time dynamics of the system are captured by the Liouville-von Neumann equation and Lindblad terms~\cite{Gerry2004}:
\begin{equation}
\frac{\partial}{\partial t}\hat{\rho} = -\frac{i}{\hbar}[ \hat{H} ,\hat{\rho}] + \frac{1}{2}\left(2\hat{L}\hat{\rho}\hat{L}^\dag-\hat{\rho}\hat{L}^\dag \hat{L}-\hat{L}^\dag \hat{L}\hat{\rho} \right).
    \label{eq:vN}
\end{equation}
\indent For the RF scans, we numerically solve the master equations for the excited-state population, $\rho_\mathrm{ee}$. We add the inhomogenous broadening via a Lorentzian probability distribution in the laser detuning to describe the frequency jitter. The full-width-at-half-maximum linewidth is \qty{300}{\MHz}. Figure~\ref{fig:3s}(a) and (b) show a comparison of the mechanically modulated RF scan, with and without inhomogeneous broadening in the simulation. As discussed in the main text, with only homogeneous broadening, clearly resolved side peaks appear but the side peaks are smeared out by the inhomogeneous broadening. \\
\indent To simulate the autocorrelation function and the emission spectrum, we make use of the quantum simulation toolbox Qutip~\cite{Johansson2012,Johansson2013}. On account of the time dependence in the Hamiltonian, we solve the correlation functions numerically. The two-time second-order correlation function reads:
\begin{equation}
    g^{(2)}(t,\tau)=\frac{\langle \hat{\sigma}_+(t)\hat{\sigma}_+(t+\tau)\hat{\sigma}_-(t+\tau)\hat{\sigma}_-(t)\rangle}{\langle \hat{\sigma}_+(t)\hat{\sigma}_-(t)\rangle^2}. 
    \label{eq:FB_two_time_corr}
\end{equation}
We solve for $t = 50\cdot T_\mathrm{m}$ and average the result over the final 20 mechanical periods, from $t = 30\cdot T_\mathrm{m}$ to $t = 50\cdot T_\mathrm{m}$, where $T_\mathrm{m} = 2\pi/\Omega_\mathrm{m}$. This is to make sure that the simulations are not influenced by the start parameters. The emission spectrum is obtained by performing a Fourier transform on the two-time first-order coherence function: 
\begin{equation}
    G^{(1)}(t,\tau) =\langle \hat{\sigma}_\mathrm{+}(t)\hat{\sigma}_\mathrm{-}(t+\tau)\rangle,
\end{equation}
\begin{equation}
S(t,\omega) =\int_{0}^{\infty} G^{(1)}(t,\tau)e^{\mathrm{-i\omega\tau}} d\tau.
\end{equation}
We average the time-dependent emission spectrum from $t = 7\cdot T_\mathrm{m}$ to $t = 10\cdot T_\mathrm{m}$ and apply a Lorentzian filter function according to the specifications of the Fabry-Perot etalon (Thorlabs, SA200-8B).\\
\indent Figure~\ref{fig:3s}(c) and (d) show a comparison of the mechanically modulated autocorrelation, with and without inhomogeneous broadening in the simulation. Clearly, the ``width" of the anti-bunching dip is reduced by the inhomogeneous broadening. As expected, without inhomogeneous broadening, no time modulation would be observed in the autocorrelation with a resonant drive. Overall, using the values from the finite-element simulations and from the QD characterisation, our model reproduces the measured data very well, from the RF scans to the RF emission spectra. \\
\indent For the simulations of the emission spectra, excited-state dephasing is added, which is obtained from fitting the autocorrelation simulations, $\gamma^*=0.3\Gamma_\mathrm{R}$. Figure~\ref{fig:3s}(e) and (f) show a comparison of the phonon-sideband (PSB) and zero-phonon line (ZPL) intensity under driving of $\mathrm{m_7}$, with and without inhomogeneous broadening, respectively. As in the experiment (Fig.~\ref{fig:3}(d)), the laser is detuned to $\Delta\omega_\mathrm{l} = -\Omega_\mathrm{m_7}$. Without inhomogeneous broadening (or in the resolved-sideband regime), the anti-Stokes sideband (PSB+1) exceeds the ZPL. With inhomogeneous broadening (or in the unresolved-sideband regime), this is not the case. Equivalently, a signature of the resolved-sideband regime is that the first PSB can become more intense than the ZPL on scanning the laser detuning.  


%

\end{document}